 \documentclass[final,5p,times,twocolumn,sort&compress,nopreprintline]{elsarticle}

\usepackage[utf8]{inputenc}
\usepackage[T1]{fontenc}
\usepackage{color}
\usepackage{latexsym,amsmath,amssymb,graphicx,booktabs}
\usepackage{hyperref}

\definecolor{MyBlue}{rgb}{0.15,0.15,0.70}
\hypersetup{
colorlinks=true,
citecolor=MyBlue,
linkcolor=MyBlue,
urlcolor=MyBlue
}

\newcommand{\Mc}{{\cal M}_c}
\newcommand{\dgw}{d_L^{\,\rm gw}}
\newcommand{\dem}{d_L^{\,\rm em}}

\renewcommand\({\left(}
\renewcommand\){\right)}
\renewcommand\[{\left[}
\renewcommand\]{\right]}

\newcommand{\ra}{\rightarrow}

\def\lsim{\raise 0.4ex\hbox{$<$}\kern -0.8em\lower 0.62
ex\hbox{$\sim$}}

\def\gsim{\raise 0.4ex\hbox{$>$}\kern -0.7em\lower 0.62
ex\hbox{$\sim$}}

\def\lbar{{\hbox{$\lambda$}\kern -0.7em\raise 0.6ex
\hbox{$-$}}}

\newcommand\eq[1]{eq.~(\ref{#1})}

\newcommand\pa{\partial}
\newcommand\p{\partial}

\newcommand\ee{\end{equation}}
\newcommand\be{\begin{equation}}
\def\bea{\begin{array}}
\def\eea{\end{array}}\def\ea{\end{array}}
\newcommand\ees{\end{eqnarray}}
\newcommand\bees{\begin{eqnarray}}

\def\d{\delta}

\def\dslash{\hspace{-1mm}\not{\hbox{\kern-2pt $\partial$}}}
\def\Dslash{\not{\hbox{\kern-2pt $D$}}}
\def\pslash{\not{\hbox{\kern-2.1pt $p$}}}
\def\kslash{\not{\hbox{\kern-2.3pt $k$}}}
\def\qslash{\not{\hbox{\kern-2.3pt $q$}}}

\def\p1{{\bf p}_1}
\def\p2{{\bf p}_2}
\def\k1{{\bf k}_1}
\def\k2{{\bf k}_2}

\newcommand{\dddM}{\kern 0.2em \raise 1.9ex\hbox{$...$}\kern -1.0em \hbox{$M$}}
\newcommand{\dddQ}{\kern 0.2em \raise 1.9ex\hbox{$...$}\kern -1.0em \hbox{$Q$}}
\newcommand{\dddI}{\kern 0.2em \raise 1.9ex\hbox{$...$}\kern -1.0em\hbox{$I$}}
\newcommand{\dddJ}{\kern 0.2em \raise 1.9ex\hbox{$...$}\kern-1.0em
\hbox{$J$}}
\newcommand{\dddcalJ}{\kern 0.2em \raise 1.9ex\hbox{$...$}\kern-1.0em
\hbox{${\cal J}$}}

\newcommand{\dddO}{\kern 0.2em \raise 1.9ex\hbox{$...$}\kern -1.0em
\hbox{${\cal O}$}}
\def\dddz{\raise 1.5ex\hbox{$...$}\kern -0.8em \hbox{$z$}}
\def\dddd{\raise 1.8ex\hbox{$...$}\kern -0.8em \hbox{$d$}}
\def\dddbd{\raise 1.8ex\hbox{$...$}\kern -0.8em \hbox{${\bf d}$}}
\def\ddbd{\raise 1.8ex\hbox{$..$}\kern -0.8em \hbox{${\bf d}$}}
\def\dddx{\raise 1.6ex\hbox{$...$}\kern -0.8em \hbox{$x$}}

\newcommand{\msun}{M_{\odot}}

\newcommand{\oma}{\Omega_{M}}

\graphicspath{ {./Graphs/} }

\begin{document}

\begin{frontmatter}



\title{Modified gravitational wave propagation and the binary neutron star mass function}


%

\author[1]{Andreas Finke}\ead{andreas.finke@unige.ch}
\author[1]{Stefano Foffa}\ead{stefano.foffa@unige.ch}
\author[1]{Francesco Iacovelli}\ead{francesco.iacovelli@unige.ch}
\author[1]{Michele Maggiore\corref{cor1}}\ead{michele.maggiore@unige.ch}
\author[1]{Michele Mancarella}\ead{michele.mancarella@unige.ch}

\cortext[cor1]{Corresponding author}

\affiliation[1]{organization={D\'epartement de Physique Th\'eorique and Center for Astroparticle Physics, Universit\'e de Gen\`eve},
            addressline={24 quai Ansermet}, 
            postcode={CH--1211 Gen\`eve 4}, 
            country={Switzerland}}
            

\begin{abstract}
Modified gravitational wave (GW) propagation is a generic phenomenon  in modified gravity. It affects the reconstruction of the redshift of  coalescing binaries from the  luminosity distance measured by GW detectors, and  therefore  the reconstruction of the actual masses of the component compact stars from the observed (`detector-frame') masses. We show that, thanks to the narrowness of the mass distribution of binary neutron stars, this effect can provide a clear signature of modified gravity, particularly for the redshifts explored by third generation GW detectors such as Einstein Telescope and Cosmic Explorer.
\end{abstract}




\end{frontmatter}





\section{Introduction}

In recent years, modified gravitational wave (GW) propagation has come to attention as one of the most promising ways of testing deviations from General Relativity (GR) on cosmological scales. The effect is encoded in the propagation equation  of  GWs across cosmological distances which,  in  modified gravity theories, can take the form~\cite{Saltas:2014dha,Lombriser:2015sxa,Nishizawa:2017nef,Arai:2017hxj,Belgacem:2017ihm,Amendola:2017ovw,Belgacem:2018lbp,Belgacem:2019pkk}
\be\label{prophmodgrav}
\tilde{h}''_A  +2 {\cal H}[1-\delta(\eta)] \tilde{h}'_A+c^2k^2\tilde{h}_A=0\, ,
\ee
where  
$\tilde{h}_A(\eta;k)$ is the Fourier transform of the GW perturbation,
 $h'=\pa h/\pa \eta$ where $\eta$ is conformal time, $a(\eta)$ is the  scale factor,   ${\cal H}=a'/a$, and $A={+,\times}$ labels the two polarizations.

The difference with respect to GR is given by a non-vanishing function $\delta(\eta)$. Several other modifications with respect to GR are possible in the propagation equation of GWs. The most immediate  options are  a deviation of the speed of  GWs from the speed of light, or a graviton mass. Both would rather modify the $c^2k^2$ term in \eq{prophmodgrav}, but are now very significantly constrained:
a deviation of the speed of  GWs from the speed of light is excluded at the level  $|c_{\rm gw}-c|/c< {\cal O}(10^{-15})$ by the observation of GW170817 and its electromagnetic counterpart \cite{Monitor:2017mdv} 
(and a large class of  modifications of GR have been ruled out by this limit~\cite{Creminelli:2017sry,Sakstein:2017xjx,Ezquiaga:2017ekz,Baker:2017hug}), 
while limits on the graviton mass are in the range   ${\cal O}(10^{-32}-10^{-22})$~eV, depending on the  probes used
\cite{ParticleDataGroup:2018ovx}; several other modifications,  in general related to rather  specific classes of modified gravity  theories, have been tested or proposed, such as extra polarizations~\cite{Will:2014kxa}, Lorentz-violating dispersion relations~\cite{Mirshekari:2011yq}, parity-violating effects~\cite{Crowder:2012ik},  or  scale dependent modifications of the speed of GWs~\cite{deRham:2018red}. 

Modified GW propagation, in the form described by \eq{prophmodgrav}, was first found in some explicit scalar-tensor theories of the Horndeski class~\cite{Saltas:2014dha,Lombriser:2015sxa,Nishizawa:2017nef,Arai:2017hxj} (see also \cite{Gleyzes:2014rba} for a discussion  within  the  effective field theory approach to dark energy) and, in  refs.~\cite{Belgacem:2017ihm,Belgacem:2018lbp}, in non-local infrared 
modifications of gravity, i.e. in theories where the  underlying classical action is still GR, but non-local terms, relevant in the infrared, are assumed to be generated by non-perturbative effects in the quantum effective action~\cite{Maggiore:2013mea} (see \cite{Belgacem:2020pdz} for recent review). However,   it has been understood that the phenomenon is completely general and appears in all best studied modified gravity theories~\cite{Belgacem:2019pkk}. It also appears, in a different form not described by \eq{prophmodgrav},  in theories with extra dimensions, where it is rather due to the loss of gravitons to the bulk~\cite{Deffayet:2007kf,Pardo:2018ipy}.

The modified friction term in \eq{prophmodgrav} changes the evolution of the GW amplitude in the  propagation across cosmological distances. Since, in GR, the amplitude of a coalescing binary is proportional to $1/d_L$, where $d_L$ is the luminosity distance, this introduces a bias in the luminosity distance  inferred from GW observations. In particular, if $\delta(\eta)<0$, the damping term is stronger and, after propagation from the source to the detector,  the GW has a smaller amplitude. If interpreted within GR,  it would therefore appear to come from a  distance larger than its actual distance (and vice versa for $\delta(\eta)>0$).
It is then useful to introduce a distinction between the standard luminosity distance, that, in this context, is called the `electromagnetic luminosity distance' and denoted by $\dem$, and the luminosity distance extracted from the observation of the GWs of a compact binary coalescence, that  is called 
the `GW luminosity distance'~\cite{Belgacem:2017ihm} and denoted by
 $\dgw$. The two quantities are related  by~\cite{Belgacem:2017ihm,Belgacem:2018lbp}
\be\label{dLgwdLem}
\dgw(z)=\dem(z)\exp\left\{-\int_0^z \,\frac{dz'}{1+z'}\,\delta(z')\right\}\, ,
\ee
where   $\delta(z)\equiv \delta[\eta(z)]$.
A useful parametrization of this effect, which catches  the  redshift dependence predicted by almost all explicit models  in terms of just two parameters $(\Xi_0,n)$,  is obtained writing~\cite{Belgacem:2018lbp},
\be\label{eq:fit}
\frac{d_L^{\,\rm gw}(z)}{d_L^{\,\rm em}(z)}=\Xi_0 +\frac{1-\Xi_0}{(1+z)^n} \, ,
\ee
which interpolates between $\dgw/\dem=1$ as $z\ra 0$ and an asymptotic value $\Xi_0$ at large $z$, with a power-law behavior in $a=1/(1+z)$ fixed by $n$.
GR is recovered  when $\Xi_0=1$ (for all $n$). The study of explicit modified gravity models shows that $\Xi_0$ can be significantly different from $1$. In particular, in non-local gravity   it can be as large  as $1.80$
\cite{Belgacem:2019lwx,Belgacem:2020pdz}, corresponding to a $80\%$ deviation from  GR, despite the fact that this model complies with existing observational bounds, that force deviations from GR and from $\Lambda$CDM in the  background evolution and  in the scalar perturbation sector to be at most of  a few percent~\cite{Aghanim:2018eyx,Abbott:2018xao}. Thus, the newly opened window of GWs could give us the best opportunities  for testing modified gravity and dark energy.

Contrary to quantities such as the speed of GWs or the graviton mass, the limits on the parameter $\Xi_0$ (the main parameter that describes modified GW propagation; the power $n$ in \eq{eq:fit} only determines the precise form of the interpolation between the asymptotic values) are still quite broad. Using the binary neutron star (BNS) GW170817, with the redshift determined from the  electromagnetic counterpart, only gives bounds of order $\Xi_0\, \lsim \, 14$  ($68\%$ c.l.)~\cite{Belgacem:2018lbp} (see also \cite{Arai:2017hxj,Lagos:2019kds}). This is because the redshift of GW170817 is very small, $z\simeq 0.01$, and $\dgw(z)/\dem(z)$ goes to one as $z\ra 0$, for all $\Xi_0$.
A more significant  limit, 
\be\label{Xi0GLADE}
\Xi_0=2.1^{+3.2}_{-1.2}\, , \qquad (68\%\, {\rm c.l.})\, ,
\ee  
has been obtained in \cite{Finke:2021aom}, using  binary black hole (BBH) coalescences without electromagnetic counterpart (`dark sirens') from the O1, O2 and O3a  runs of the  LIGO/Virgo Collaboration (LVC) and correlating them with the GLADE galaxy catalog~\cite{Dalya:2018cnd}.  An even more stringent measurement  is obtained under the tentative  identification of the flare  ZTF19abanrhr as the electromagnetic counterpart of the BBH coalescence GW190521, in which case
one gets $\Xi_0=1.8^{+0.9}_{-0.6}$   \cite{Finke:2021aom}    (see also \cite{Mastrogiovanni:2020mvm}). However, this identification currently is not secure. A   limit on modified GW propagation (using a different parametrization) has been obtained in \cite{Ezquiaga:2021ayr}  using the BBH mass function,  following an idea originally proposed in \cite{Farr:2019twy} to infer $H_0$, and a recent re-analysis in~\cite{Mancarella:2021ecn}, using again the BBH mass function, gives 
\be\label{Xi0fromBBH}
\Xi_0=1.2\pm 0.7 \, , \qquad (68\%\, {\rm c.l.})\, ,
\ee 
while the corresponding  limit at $90\%$ c.l. is 
\be\label{Xi0fromBBH90}
\Xi_0=1.2^{+1.5}_{-1.0} \, , \qquad (90\%\,  {\rm c.l.})\, .
\ee 
Even with the study based on the BBH mass function, which currently gives the most stringent bounds on $\Xi_0$, current data are not constraining enough to obtain a limit on $n$, with the posterior reflecting basically the prior used~\cite{Mancarella:2021ecn}.

Since the effect of modified GW propagation increases with redshift (at least until the ratio in \eq{eq:fit}
saturates to its large $z$ limit $\dgw/\dem\simeq \Xi_0$), third generation (3G) ground based GW detectors such the  Einstein Telescope (ET)~\cite{Punturo:2010zz,Maggiore:2019uih} and Cosmic Explorer (CE)~\cite{Reitze:2019iox}, or   the  space interferometer LISA~\cite{Audley:2017drz}, are particularly well suited to study it, and several forecasts have been made on the accuracy that future observations can reach on $\Xi_0$, using different techniques~\cite{Belgacem:2019tbw,Belgacem:2019pkk,Belgacem:2019zzu,Mastrogiovanni:2020gua,Baker:2020apq,Mukherjee:2020mha,Ye:2021klk,Jiang:2021mpd,Canas-Herrera:2021qxs,Finke:2021znb}.  Observe that the function $\delta(\eta)$ in \eq{prophmodgrav} only affects the amplitude of the GW signal. Other effects, such as modified dispersion relations, also affect the  post-Newtonian coefficients of the phase (see \cite{LIGOScientific:2021sio} for the most recent bounds using LIGO/Virgo data). In that case, eventually, a joint analysis of the Hubble parameter $H_0$, of modified GW propagation and of modified dispersion relations might be necessary~\cite{Mastrogiovanni:2020gua}.\footnote{Note, however,  that a modified gravity theory such as the nonlocal gravity model mentioned above predicts that the only modification with respect to GR will be given by the function $\delta(\eta)$ in \eq{prophmodgrav}, and fits the cosmological data with a value of $H_0$ very close to that of $\Lambda$CDM.}

The same logic that has been used in \cite{Ezquiaga:2021ayr,Mancarella:2021ecn} to obtain bounds on modified GW propagation from the BBH mass function can be applied to the BNS mass function, with the advantage that the latter is narrow and is not expected to evolve significantly with redshift. 
The idea of using the BNS mass function for extracting cosmological informations was proposed in 
\cite{Chernoff:1993th,Taylor:2011fs,Taylor:2012db} in the context of the determination of $H_0$ within $\Lambda$CDM. In this paper we will discuss its application to modified GW propagation.
We will see that the method based on the BNS mass function can become particularly powerful when applied to modified GW propagation at 3G detectors, thanks to the fact that modified GW propagation increases with distance, and ET and CE can detect BNS up to  large redshifts, $z\simeq 2-3$ for ET, and even $z\sim 10$ for CE~\cite{Hall:2019xmm}.\footnote{In contrast, LISA is not sensitive to BNS mergers, nor to BNS inspirals at cosmological distances, so this method only applies to 3G ground-based detectors.}

\section{Modified GW propagation and mass reconstruction}

\begin{figure}[t]
\centering
\includegraphics[width=0.5\textwidth]{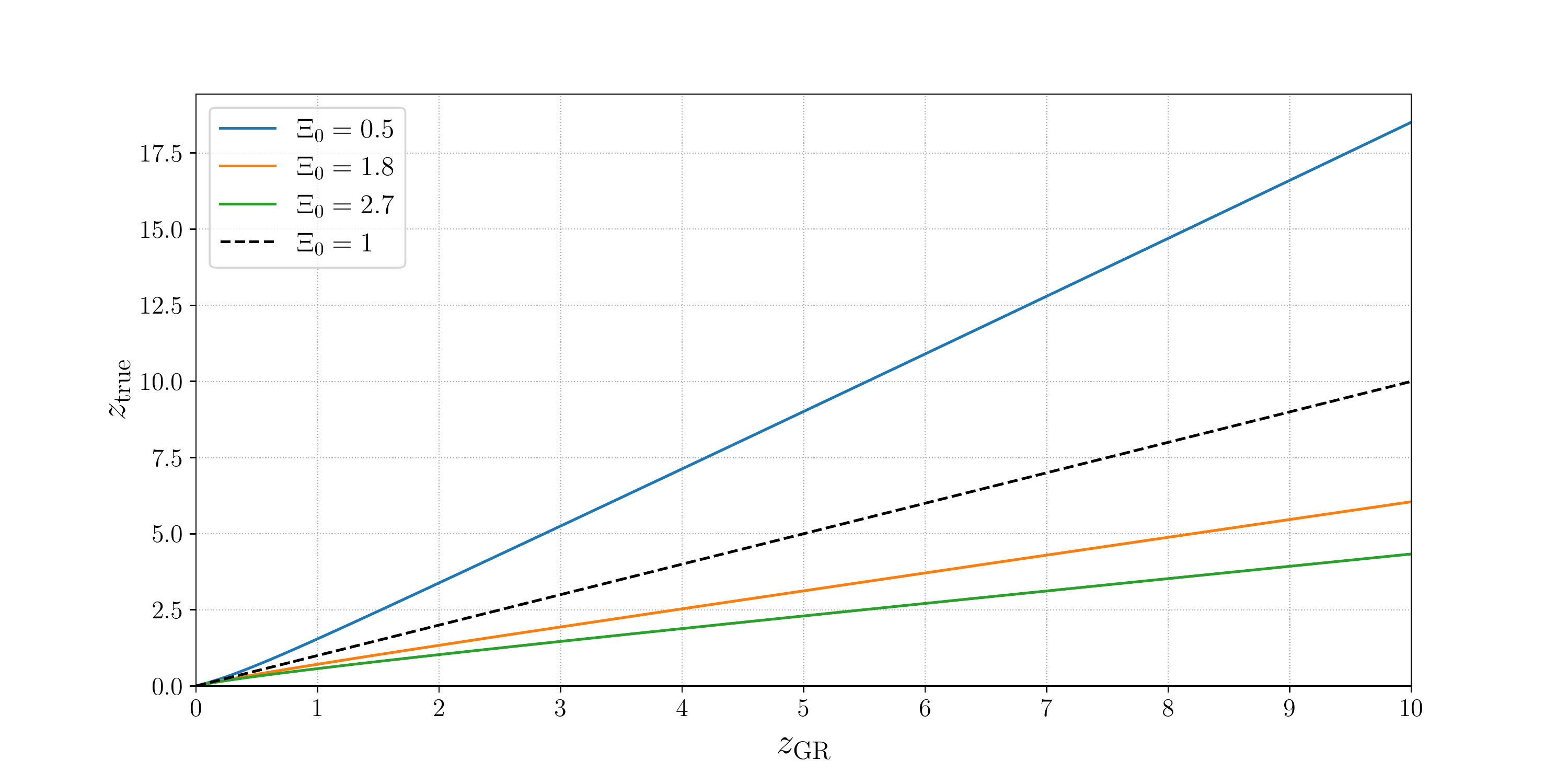}
\caption{The  redshift $z_{\rm true}$ of a source, as a function of the value $z_{\rm GR}$ that would be incorrectly inferred using GR
if Nature is  described by a modified  gravity  theory with $\Xi_0\neq 1$, for different values of $\Xi_0$
[we  set for definiteness $n=1.9$ in \eq{eq:fit}, which is a value suggested by non-local gravity, but the precise value of  $n$ has little effect.]}
\label{fig:zgw_from_zem}
\end{figure}

The starting point of our analysis is the fact that GW detectors measure the GW luminosity distance of the source, $\dgw$, which is different from the actual electromagnetic luminosity distance $\dem$ if, in Nature, $\Xi_0\neq 1$. The redshift $z_{\rm GR}$ of the source  inferred from the measured $\dgw$ assuming GR and $\Lambda$CDM (with a given value of $H_0$ and $\oma$, that we keep for definiteness the same in GR and in the modified gravity theory under consideration), would therefore differ from the true value $z_{\rm true}$. The effect is shown in Fig.~\ref{fig:zgw_from_zem}, for a sample of values of $\Xi_0$ consistent with \eq{Xi0fromBBH90}. We see that the effect can become very significant at large redshifts.

\begin{figure}[t]
\centering
\includegraphics[width=0.5\textwidth]{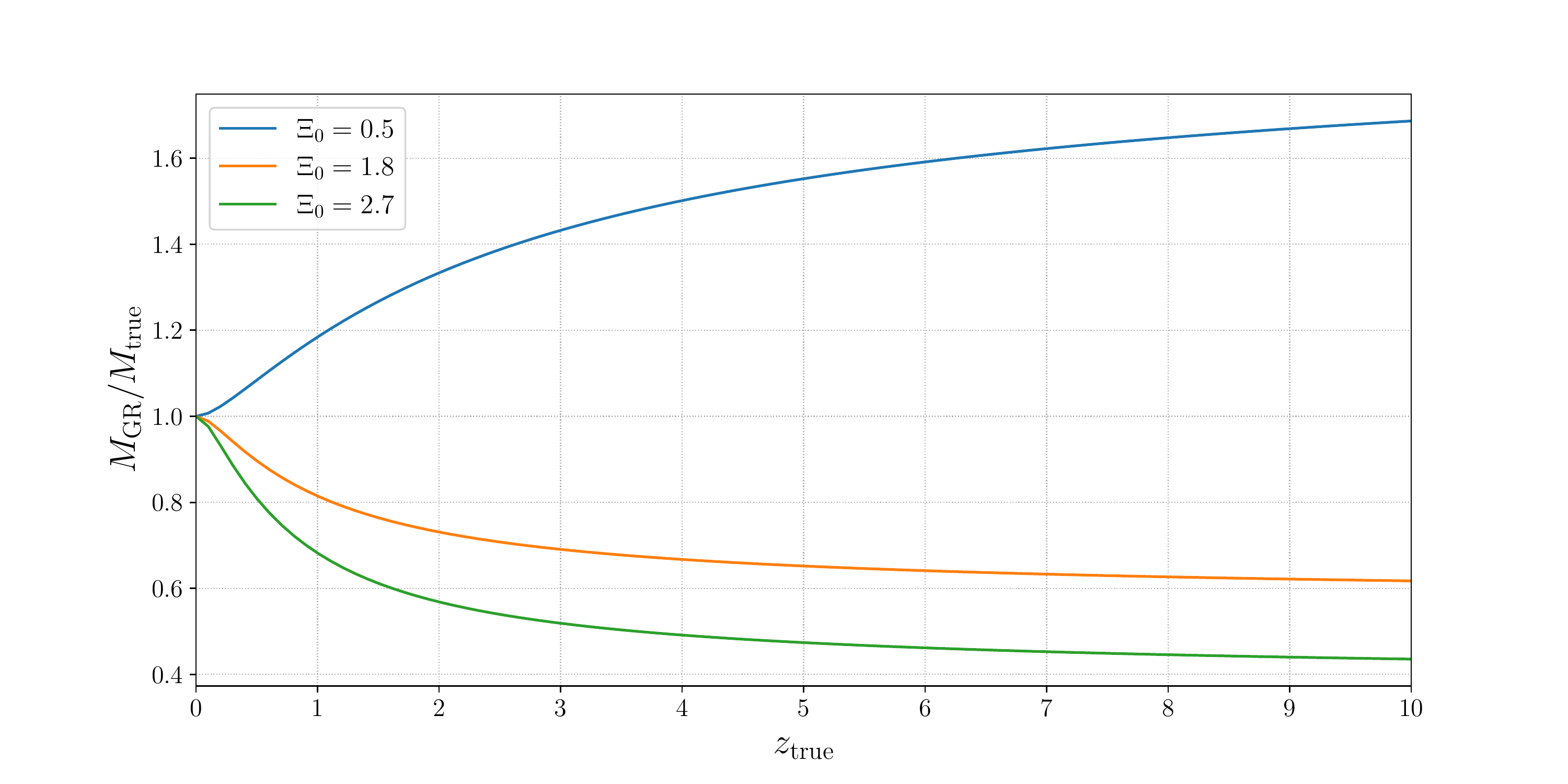}
\includegraphics[width=0.5\textwidth]{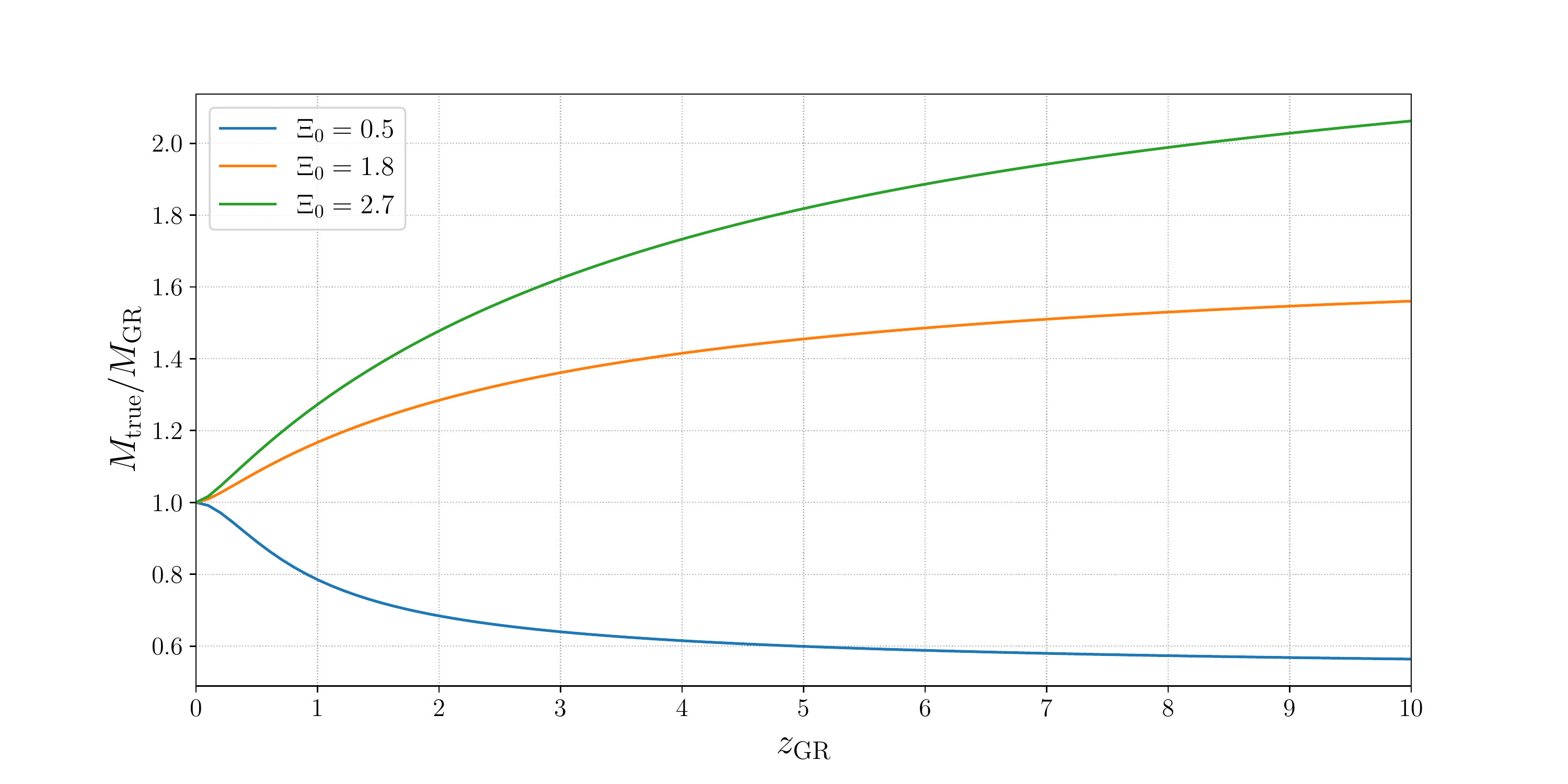}
\caption{Upper panel: the  ratio  of the mass $M_{\rm GR}$ inferred in GR over  the true mass $M_{\rm true}$, as a function of the redshift $z_{\rm true}$. Lower panel:
the  ratio  of the true mass $M_{\rm true}$ over the mass $M_{\rm GR}$ inferred in GR, as a function of the redshift $z_{\rm GR}$  inferred in GR. Here $M$ represent any mass scale made with the masses of the component stars, such as individual component masses, total mass, or chirp mass.}
\label{fig:mGR_over_mtrue}
\end{figure}

In turn, this affects the reconstruction of the  actual  masses $m_i$ ($i=1,2$) of the component stars (`source-frame' masses, as they are called in this context),  from the  `detector-frame' masses  $m_{{\rm (\rm det)}, i}\equiv  (1+z) m_i$, that are the quantities directly obtained from the GW observations. 
If Nature is described by a modified gravity theory with $\Xi_0\neq 1$, the true values  of the source-frame masses, $m_{{\rm true}, i}$, are related to the values  of the source-frame masses that would be inferred in GR, $m_{{\rm GR},i}$, by
\be\label{msource_true}
m_{{\rm true}, i}=\frac{ m_{{\rm (\rm det)}, i} }{1+z_{\rm true} } = 
\(\frac{1+z_{\rm GR} }{ 1+z_{\rm true} }\)\,  m_{{\rm GR},i}\, ,
\ee
where $m_{{\rm GR},i}  = m_{{\rm (\rm det)}, i}/(1+z_{\rm GR})$. 
The same multiplicative bias factor   will appear in  any other combination with  dimensions of mass  of the source-frame masses of the component stars, such as the total source-frame mass $m_{\rm tot}=m_1+m_2$, or the source-frame chirp mass $M_c=(m_1m_2)^{3/5}/m_{\rm tot}^{1/5}$. 
The upper panel of Fig.~\ref{fig:mGR_over_mtrue} shows the   ratio   $M_{\rm GR}/M_{\rm true}$, as a function of $z_{\rm true}$ (while the lower panel shows 
$M_{\rm true}/M_{\rm GR}$, as a function of $z_{\rm GR}$), for any such mass scale. 
We see that, at the redshifts accessible to ET and CE, and for values of $\Xi_0$ consistent with current limits, 
the effect can be very large. For instance, setting $\Xi_0=1.8$, for a  NS with $m_{\rm true}=1.35\msun$ at $z_{\rm true}=1$ (that, with this value of $\Xi_0$, corresponds to $z_{\rm GR}\simeq 1.45$),  the   mass incorrectly inferred from GR would be  $m_{\rm GR}\simeq 1.10\msun $; at $z_{\rm true}=2$ ($z_{\rm GR}\simeq 3.10$) for the same system in GR one would infer $m_{\rm GR}\simeq 0.99\msun $; and,
 for a BNS with the same mass at  $z_{\rm true}=5$  ($z_{\rm GR}\simeq 8.20$), which could still be  accessible to CE,   one would find  $m_{\rm GR}\simeq 0.88\msun$. Furthermore,  exactly the same factor affects the two component stars (which is not the case in general for  astrophysical  effects), so  a BNS with $(1.35+1.35)\msun$ would appear as a  $(1.10+1.10)\msun$ system for $z_{\rm true}=1$,
as a  $(0.99+0.99)\msun$ system for $z_{\rm true}=2$, 
and as a  $(0.88+0.88)\msun$ system at  $z_{\rm true}=5$.
 Compared to the narrowness of the neutron star (NS) mass distribution, this is a huge effect. 
The  mass of  the BNSs found  with electromagnetic observations  can be described by a Gaussian distribution with mean $1.33 \msun$ and standard deviation $0.09 \msun$~\cite{Farrow:2019xnc} (which, assuming that the distribution of the two masses are independent, corresponds to a Gaussian distribution for the total mass with mean
$2.66 \msun$ and standard deviation $0.13 \msun$), or by  a flat distribution between a minimum and a maximum mass, with a similar width. Somewhat broader limits are obtained  from an analysis using only the NSs in BNS or in BH-NS systems detected by  GW observations~\cite{Landry:2021hvl}, although this sample, of six NSs,  is very limited.

\begin{figure}[t]
\centering
\includegraphics[width=0.5\textwidth]{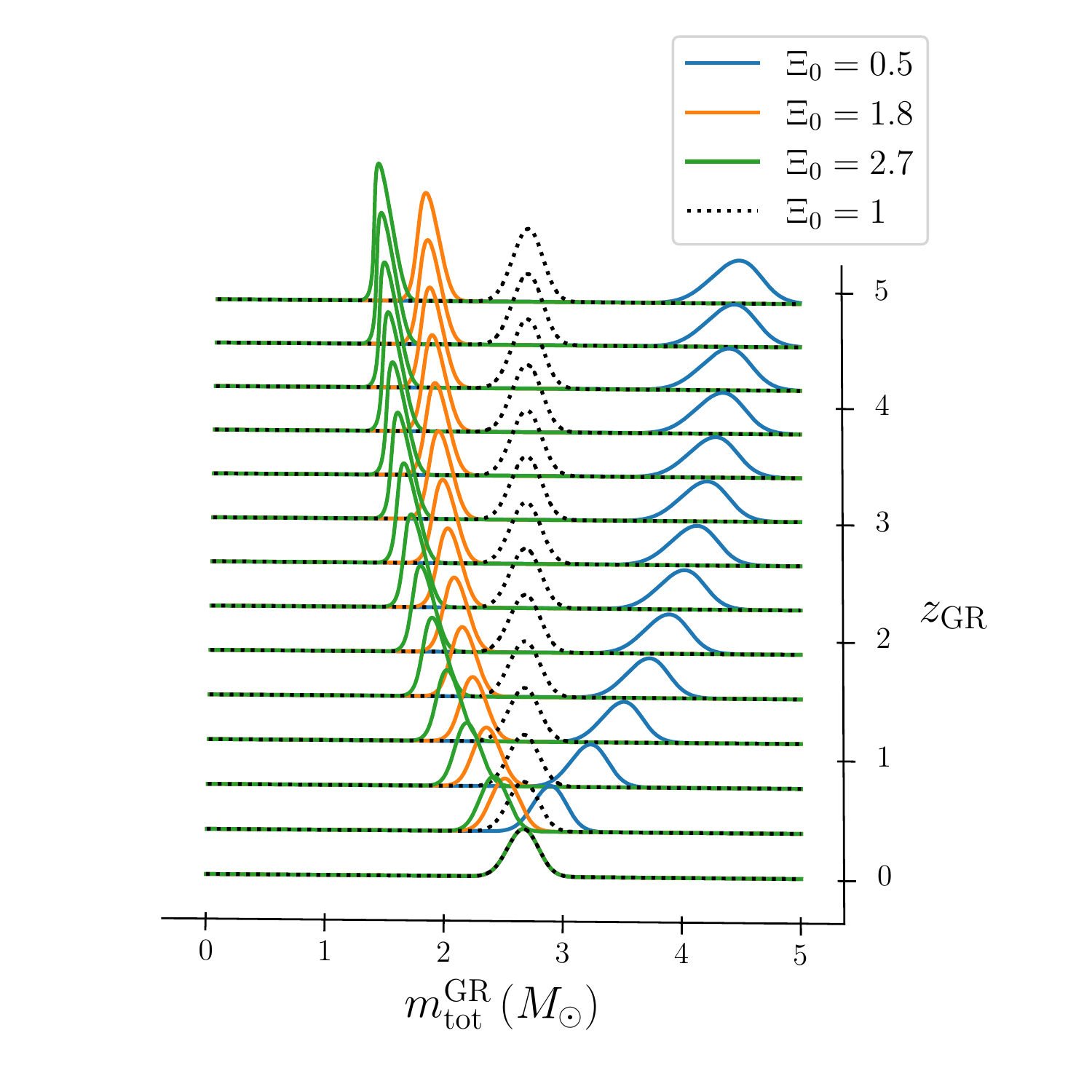}
\caption{The  evolution in redshift of the distribution of the BNS (source-frame) total mass, inferred using GR, for different values of 
$\Xi_0$. We assume a Gaussian distribution.}
\label{fig:PSR_MassDist_variousXi0_zGR}
\end{figure}

%

These estimates show that even a single BNS at large $z$ would have a significant constraining power on $\Xi_0$. Still, if one would find just a single system that, interpreted within GR, corresponds to, say, a 
$(1.0+1.0)\msun$ binary at $z_{\rm GR}\simeq 3.1$, as in one of the examples above, one would remain in doubt on whether this is a binary made of   exotic 
compact objects, such as  primordial black holes, or a signal of modified GW propagation. The power of the method, however, is that the same effect will affect {\em all} BNS systems, by a factor that depends only on $z$. If Nature is described by a modified gravity theory with a large deviation from GR such as, say, $\Xi_0=1.8$, as in the examples above, at large redshifts ET and CE will not find a single BNS whose component masses, interpreting the data within GR, will be near the typical value of  $1.35\msun$. When interpreted within GR, all BNS with $z_{\rm true}=1$  would appear to have component masses around $1.10\msun$; all BNS  at $z_{\rm true}=2$  would appear to have 
masses around $0.99\msun $, and so on.  The  detection rate of BNS at ET and CE will be impressive, of order of
$7\times 10^4$ events per year already for a single detector such as ET~\cite{Regimbau:2012ir,Regimbau:2014uia,Belgacem:2019tbw} and, among these,  within a GR interpretation, there would not be a single `normal' neutron star at large $z$, but rather a plethora of objects with puzzling masses.  The situation is illustrated in Fig.~\ref{fig:PSR_MassDist_variousXi0_zGR}, where $m_{\rm tot}^{\rm GR}$ denotes the total (source-frame) mass of the BNS inferred in GR, for different values of $\Xi_0$. Here we have assumed that the distribution of the   source-frame total mass of the binary is a Gaussian, with mean $2.66 \msun$ and standard deviation $0.13 \msun$. In the absence of astrophysical evolutionary effects, for which, currently, there is little observational information, but which are not expected by any means to give effects comparable to those shown in the figure (see, e.g., Fig.~4 of ref.~\cite{Galaudage:2020zst}), the  distribution would not change with redshift (black dotted line). In the presence of modified GW propagation, with the values of $\Xi_0$ shown in the figure, that represent  deviations from GR large but still consistent with current limits,
the masses wrongly inferred using GR are  narrowly distributed around completely different mean values. Note that the apparent skewness of the Gaussians at large $z$ is just a graphical effect  in this plot (selection effects could, however,  introduce some actual skewness, since the low mass end will be less detected at high redshifts).

\begin{figure}[t]
\centering
\includegraphics[width=0.5\textwidth]{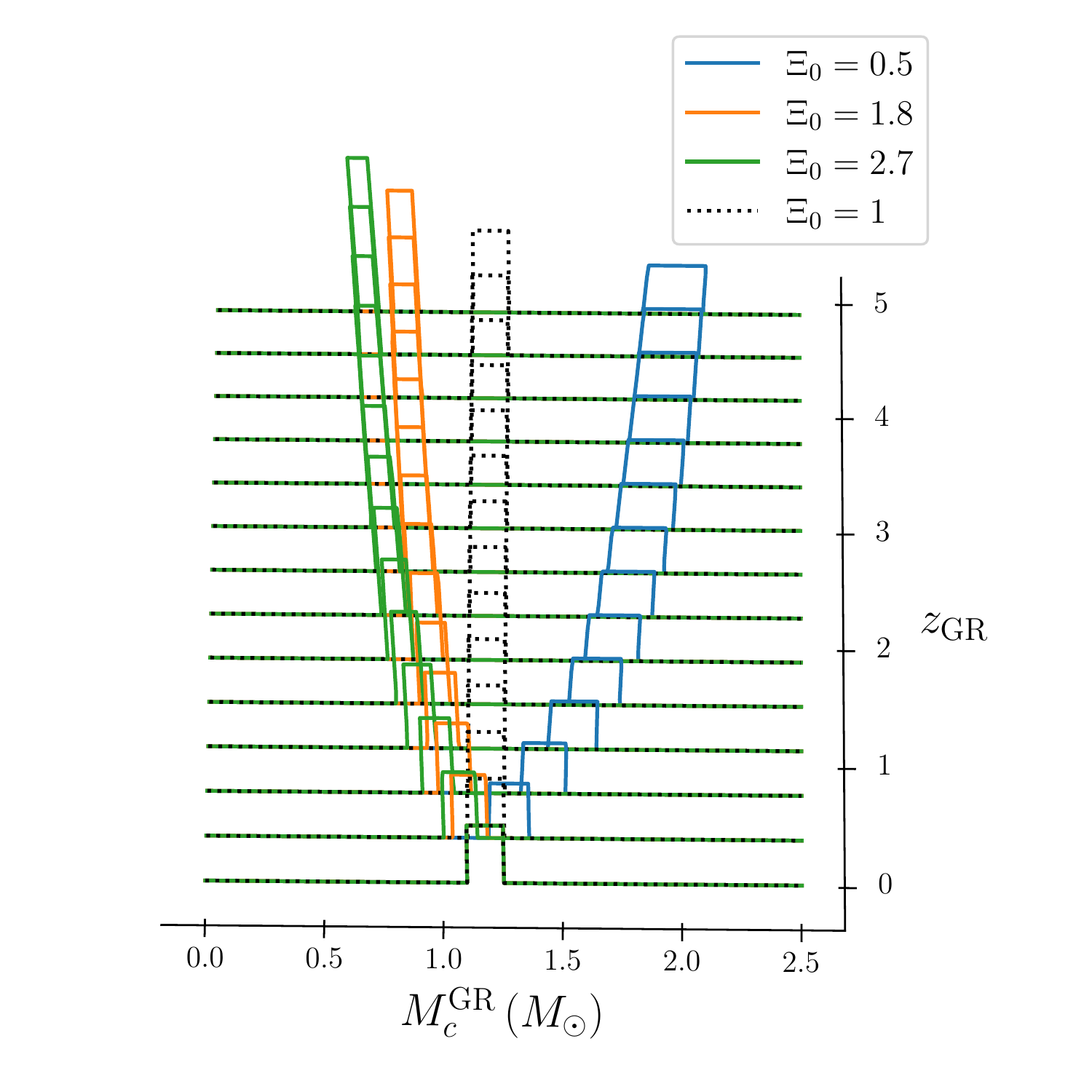}
\caption{The  evolution in redshift of the distribution of the BNS source-frame chirp mass, inferred using GR, for different values of 
$\Xi_0$. We assume a flat distribution between $M_{c,{\rm min}}\simeq 1.10\msun$ and $M_{c,{\rm max}}\simeq 1.25\msun$.}
\label{fig:PSR_MassDist_variousXi0_McFlat}
\end{figure}

Actually, it is convenient to use the chirp mass, rather than the total mass, because, in GW observations, the chirp mass is measured much more accurately than the individual component masses or the 
total mass. 
The corresponding  apparent evolution in redshift  is shown in 
Fig.~\ref{fig:PSR_MassDist_variousXi0_McFlat}. Actually, the distribution of total masses, and also of chirp masses, can be fitted both by a Gaussian distribution of by a flat distribution between minimum and maximum values. In Fig.~\ref{fig:PSR_MassDist_variousXi0_McFlat} we have used, for illustration, a flat distribution between $M_{c,{\rm min}}=1.10\msun$ and $M_{c,{\rm max}}=1.25\msun$, that encompasses the chirp masses of all BNS that merge within a Hubble time, reported in Table~1 of ~\cite{Farrow:2019xnc}.


Finally, another important signature of modified GW propagation will be given by how the BNS  population is distributed in redshift (i.e., the absolute normalization of the distributions, that  in Fig.~\ref{fig:PSR_MassDist_variousXi0_zGR} and \ref{fig:PSR_MassDist_variousXi0_McFlat} have been normalized to unity). Even if our prior information on the BNS merger rate is not as stringent as on the BNS mass function, still we expect that the rate will be described by a Madau-Dickinson form~\cite{Madau:2014bja,Madau:2016jbv,Callister:2020arv}
\be
R(z) = R_0 C_0 \frac{(1+z)^{\alpha_z}}{1 +\(  \frac{1+z}{1+z_p}   \)^{\alpha_z + \beta_z}}\, ,
\ee
where $C_0(z_p, \alpha_z, \beta_z) \equiv  1 + (1+z_p)^{-\alpha_z-\beta_z}$ is a  normalization constant 
that 
ensures  $R(0)=R_0$, and 
$z_p$ is the peak of the star formation rate, which is known to be in the range 
$z_p\simeq (2-3)$. 
In a modified gravity theory, the difference between $z_{\rm GR}$ and $z_{\rm true}$ will lead to a bias in the reconstruction of $R(z)$.\footnote{In the context  of the analysis made with the BBH mass function, this has been shown explicitly with a full Bayesian analysis in~\cite{Mancarella:2021ecn}, see, in particular, Fig.~7 of that paper.}  For instance we have seen that, if Nature is described by  a modified gravity theory with  our reference value $\Xi_0=1.8$,  and we rather use  GR to interpret the data, a BNS with $z_{\rm true}=2$  would be wrongly interpreted as having a redshift 
$z_{\rm GR}\simeq 3.10$, and  $z_{\rm true}=3$ corresponds to  $z_{\rm GR}\simeq 4.79$. The peak of the BNS merger distribution would then appear to be at redshifts larger than the peak of the star formation rate, leading to  another puzzling result of the GR interpretation (that, for $\Xi_0>1$,  could not be explained in terms of delay between formation and merger, since in this case one would find that the peak of the merger rate took place before the peak of the star formation rate). A joint Bayesian inference on the BNS mass function and on the  BNS rate parameters would therefore further strengthen the power of the method.

\section{Sources of errors}

The above discussion is still idealized, because it neglected the errors on the measurements.    The relative accuracy on the detector-frame chirp mass $\Mc=(1+z)M_c$  is of order $\Delta\Mc/\Mc\sim 1/{\cal N}_c$, where ${\cal N}_c$ is the number of inspiral cycles of the signal in the detector bandwidth, see, e.g. eq.~(7.187) of \cite{Maggiore:1900zz}. For a lower cutoff of the detector near 3~Hz, as in the design of ET, and the  chirp mass of a BNS, we have ${\cal N}_c\simeq 10^5$  (using eq.~(4.23) of \cite{Maggiore:1900zz}). The error on the detector-frame chirp mass is therefore negligible. More important is the error on the redshift due to the observational error on $\dgw$, which affects the reconstruction of the source-frame chirp mass. 
From $M_c=\Mc/(1+z)$ [where we set $(M_c=M_{c,\rm true}, z=z_{\rm true})$ but the same computation holds for 
$(M_c=M_{c,\rm GR}, z=z_{\rm GR})$], and the fact that the error on ${\cal M}_c$ is negligible, it follows that
\be\label{def_f}
\frac{\Delta M_c}{M_c}=\[\frac{\dgw}{(1+z)\pa\dgw/\pa z}\]\, \frac{\Delta\dgw}{\dgw}
\equiv f(z;\Xi_0) \frac{\Delta\dgw}{\dgw}\, .
\ee

%
\begin{figure}[t]
\centering
\includegraphics[width=0.5\textwidth]{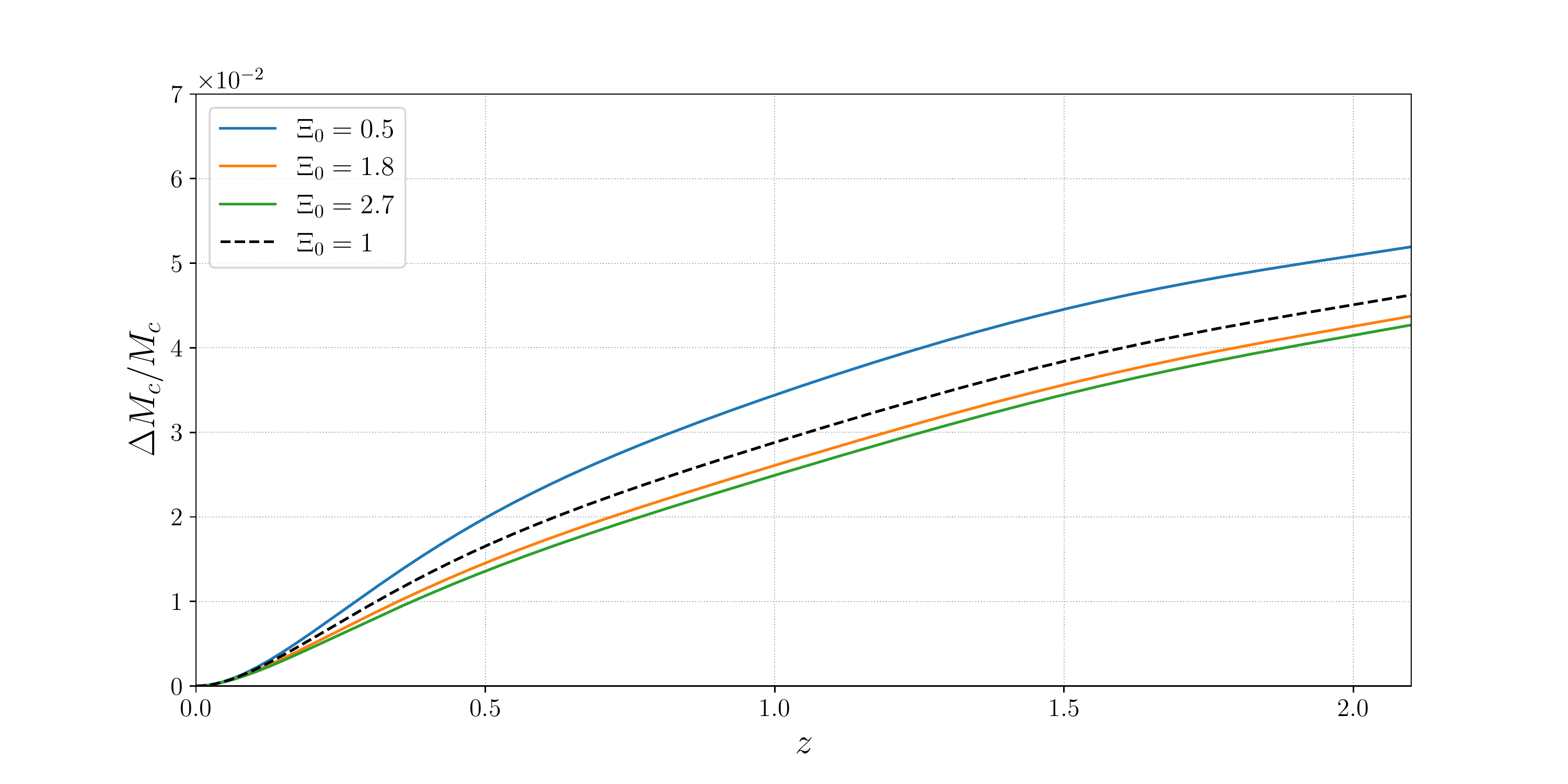}
\includegraphics[width=0.5\textwidth]{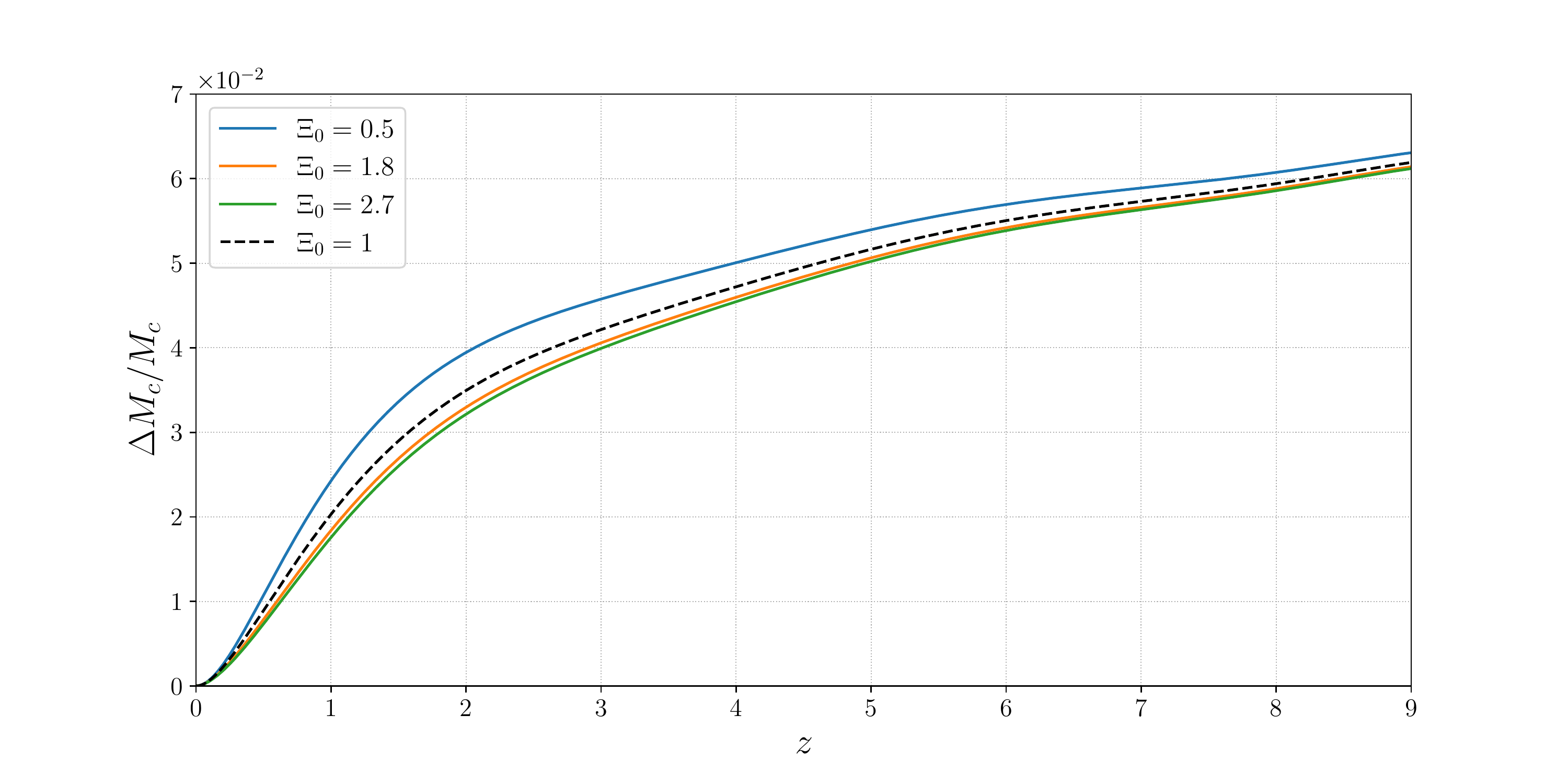}
\caption{The relative error on the the source-frame chirp mass $M_c$ due to  the observational error on $\dgw$, for different values of $\Xi_0$. Upper panel: for ET alone, over the range of redshifts available to ET. Lower panel: for a network ET+CE+CE, on a much broader redshift range.}
\label{fig:DeltaMc_over_Mc}
\end{figure}

\noindent
The function $f(z;\Xi_0)$  
goes from zero at $z=0$ to  one at large $z$, with only  mild dependence on $\Xi_0$.
The  value of 
$\Delta\dgw/\dgw$   as a function of redshift  can be obtained using   the fitting formulas provided in  \cite{Belgacem:2019tbw} [eq.~(2.13)  for ET, and eq.~(2.20)  for a network ET+CE+CE],  which were obtained  from a mock source catalog of BNS detections,  averaging   over detector orientation, source inclination, and BNS mass distribution. In Fig.~\ref{fig:DeltaMc_over_Mc} we show the result  for $\Delta M_c/M_c$ at ET (upper panel) and at a network ET+CE+CE (lower panel). We see that, on average, the relative error on the source-frame chirp mass, induced by the observational error on $\dgw$, is 
below $6\%$ up to $z\, \lsim\,9$ for a network ET+CE+CE (where ET contributes to BNS detections only up to $z\simeq 3$).  Similarly, we find that it is below $(5-6)\%$ up to $z\, \lsim\, 3$ for ET alone.
This  is smaller than the  intrinsic relative width of the BNS mass distribution, $\Delta m/m\sim 0.1$ obtained from electromagnetic observations of BNS, and therefore also of the corresponding distribution of chirp masses. So,  the 
accuracy of the method appears to be limited more by the intrinsic width of the BNS mass distribution,  than by observational errors on the reconstruction of the redshift. 

Other sources of error would require more complex dedicated studies, that are beyond the scope of this paper. One  is the  error due to lensing from  large scale structures along the line of sight.
On linear scales, inhomogeneities induce a relative error $\Delta d_L/d_L \,\lsim\, 1\%$ for all redshifts $z<5$ \cite{Bertacca:2017vod} (see also Fig.~12 of \cite{Maggiore:2019uih}).  Therefore, these are smaller than the error on the measurement of the luminosity distance in ET. The treatment of non-linear scales is, however,  more complex and has been recently discussed in~\cite{Kalomenopoulos:2020klp}. In this case, using a simplified model for modified GW propagation, corresponding to setting $\delta(z)$ equal to a constant $\d_0$ in \eq{dLgwdLem},\footnote{This is a special case of \eq{eq:fit}, obtained setting $\Xi_0=0$ and $n=\delta_0$. However, all explicit examples of modified gravity models worked out to date  rather predict a function $\delta(z)$ that goes to zero at large redshifts, corresponding to the fact that dark energy turns on in a recent cosmological epoch, and $\Xi_0$ is never close to zero, so this modelization is not very realistic.}  and using the notation $\nu=-2\delta_0$, ref.~\cite{Kalomenopoulos:2020klp} finds that, at ET, 350 BNS events with counterpart 
would be needed to measure $\nu$ at the $1\%$ level. It would be very interesting to extend the study 
in~\cite{Kalomenopoulos:2020klp} to the computation of the effect of lensing from clustered structures
on the reconstructed BNS mass function,  using furthermore the full 
expression (\ref{dLgwdLem}) for modified GW propagation.

Another important point concerns the evolution with redshift of the BNS mass function, due to astrophysical evolutionary effects. For values of $\Xi_0$ such as our reference value $\Xi_0=1.8$, that represents a  large deviations from GR, we have seen in Figs.~\ref{fig:PSR_MassDist_variousXi0_zGR} or \ref{fig:PSR_MassDist_variousXi0_McFlat}, and in the discussion around them,  that the effect of modified GW propagation is very large compared to anything that could be expected from evolutionary effects.  We certainly do not expect that at, say, redshift $z=2$, neutron stars  have a mass $m\simeq 1.0\msun$, and they only reach the typical observed  values $m\simeq 1.33\msun$ in the local Universe because of evolutionary effects; in contrast, as we have seen, $m\simeq 1.0\msun$ is the value that would be erroneously reconstructed using GR, if $\Xi_0=1.8$ and $z_{\rm true}=2$. So, at this level, evolutionary effects cannot mimic modified gravity.
However, to detect finer deviations from GR, corresponding to values of $\Xi_0$ closer to one, eventually also evolutionary effects in BNS will have to be taken into account. By the time that ET and CE will be operational, more information on the evolution of the BNS masses with redshift  might have been obtained from electromagnetic observations,  expanding the  currently very limited sample
of 17 BNS, used in \cite{Farrow:2019xnc}. Furthermore, as we already remarked, a handle to discriminate the effect of modified GW propagation from evolutionary effects is that the former acts exactly in the same way on the reconstruction of the two star masses, and therefore does not affect the inferred mass ratio, while  evolutionary effects in general modify the mass ratio.
Eventually, the best strategy will be to perform a joint inference of the cosmological parameters and of the parameters describing the astrophysical population, along the lines discussed in \cite{Farr:2019twy,Ezquiaga:2020tns,You:2020wju,Mastrogiovanni:2021wsd,Ezquiaga:2021ayr,Mancarella:2021ecn}, including 
all this prior information.

A full Bayesian analysis on mock data for  3G detectors, including selection effects and the current understanding of observational errors,  which is necessary to reliably quantify  the accuracy that can be obtained on $\Xi_0$, is under development and will be presented in a separate paper.

\section{Conclusions}

For BNSs,  at the large redshifts that will be probed by third-generation detectors such as Einstein Telescope and Cosmic Explorer, modified GW propagation could leave a very characteristic imprint on the mass distribution (and, to some extent, also on the redshift distribution)  of the observed  BNS. In modified gravity, the size of the effect is  controlled by the parameter $\Xi_0$ introduced in \eq{eq:fit}. For values as large as $\Xi_0\simeq 1.8$, that are consistent with current limits and are on the upper range of the predictions from  an explicit and viable model  of modified gravity~\cite{Maggiore:2013mea,Belgacem:2019lwx,Belgacem:2020pdz}, the 
effect on the reconstructions of the  BNS mass function and of the BNS merger rate is quite striking, and would  provide a clear and unambiguous signature of modifications of General Relativity on cosmological scales. For values of $\Xi_0$ closer to the GR value $\Xi_0=1$, disentangling the effect of modified gravity from astrophysical and cosmological effects (such as evolutionary effect in the BNS mass function or lensing by non-linear structures) will eventually become  more challenging and will require further studies.

\section*{Declaration of competing interest}
The authors declare that they have no known competing financial interests or personal relationships that could have appeared to influence the work reported in this paper.

\section*{Acknowledgments}
The work  of the authors is supported by the  Swiss National Science Foundation, grant number 200020\_191957, and  by the SwissMap National Center for Competence in Research.


\bibliographystyle{elsarticle-num}
\bibliography{myrefs}

\end{document}